\documentclass[aps,prd,print,twocolumn,nofootinbib,preprintnumbers]{revtex4-1}
\pdfoutput=1 
\usepackage{amsmath,amssymb,amsfonts}
\usepackage{graphicx}
\usepackage{booktabs,colortbl}
\usepackage{cancel,xcolor,multirow}
\usepackage[normal]{caption}
\usepackage{subcaption}
\usepackage{slashed}

\newcommand{\tev}{~\text{TeV}}
\newcommand{\gev}{~\text{GeV}}

\newcommand{\Lc}{\mathcal{L}}

\newcommand{\be}{\begin{equation}} 
\newcommand{\ee}{\end{equation}} 
\newcommand{\bea}{\begin{eqnarray}}  
\newcommand{\eea}{\end{eqnarray}}
\newcommand{\bs}{\begin{split}} 
\newcommand{\es}{\end{split}}

\newcommand{\units}[1]{~\mathrm{#1}}
\newcommand{\GeV}{~\mathrm{GeV}}
\newcommand{\mev}{~\mathrm{MeV}}

\newcommand{\invfb}{~\mathrm{fb^{-1}}}
\newcommand{\invab}{~\mathrm{ab^{-1}}}

\begin{document}


\title{ Constraining the minimal dark matter fiveplet with LHC searches}

\author{B. Ostdiek\footnote{E-mail: bostdiek@nd.edu}}
\affiliation{Department of Physics, University of Notre Dame\\225 Nieuwland Science Hall\\Notre Dame, IN 46556, U.S.A.}

 
\begin{abstract}
Adding a fermion to the standard model particle content which is a fiveplet under $SU(2)_L$ gives a dark matter candidate. The lightest state in the multiplet is neutral and automatically stable. The charged component of the multiplet is only slightly heavier and can travel a finite distance in the LHC detectors before decaying, leaving a charged track which disappears before the edge of the detector. We use the recent ATLAS and CMS searches to exclude a Majorana fiveplet with a mass up to 267 GeV. We estimate that with 3 ab$^{-1}$ of $\sqrt{s}=14$ TeV data this could be pushed to a mass of 520 GeV. These exclusions are $\sim 10\%$ greater than what is achieved for wino-like dark matter. We also discuss how the doubly charged states could be used to distinguish a disappearing track signal from that given by a triplet such as the pure wino.
\end{abstract}

\maketitle

\section{Introduction}
Weakly interacting massive particle (WIMP) dark matter provides encouragement for expecting new physics at the TeV scale beyond the motivation from the hierarchy problem. The well known WIMP coincidence allows for thermally produced WIMPs in the early universe to freeze out to the observed relic abundance if the masses are near the TeV scale and the annihilation cross sections in the early universe are weak sized. 

Supersymmetry with R parity provides a common example of this when the lightest neutralino is the stable dark matter candidate. Assuming the rest of the supersymmetric spectrum is too heavy to have an effect, pure Higgsino-like dark matter requires a dark matter mass near 1 TeV while pure wino dark matter requires a mass near 3 TeV to achieve the correct relic abundance \cite{Hisano:2006nn,Hryczuk:2010zi}. Dark matter masses less than this can be achieved through non-trivial mixings of the bino, wino, and Higgsinos (well tempering)\cite{ArkaniHamed:2006mb}\footnote{We ignore co-annihilations with sfermions or (pseudo)scalar funnels.}. The well tempered neutralino is lighter and contains more states nearby in mass than do the pure gauge eigenstates, which allows for search strategies utilizing the SM decay products when the heavier states decay down to the LSP \cite{Baer:2009bu, Giudice:2010wb, Schwaller:2013baa,Han:2014kaa, Low:2014cba, Bramante:2014dza,Calibbi:2014lga,Han:2014sya,Bramante:2014tba,Martin:2014qra,Han:2014xoa}. The small mass differences in the well tempered scenario make searches difficult as the visible decay products come out with very little energy. This problem is exacerbated in the pure gauge eigenstate scenarios where the mass splitting only comes from loop effects. Search strategies for such scenarios must rely on something other than standard model (SM) decay products of the process.

In anomaly mediated supersymmetry breaking (AMSB) scenarios \cite{Randall:1998uk, Giudice:1998xp, Moroi:1999zb}, the LSP is the neutral wino which is nearly degenerate with the charged wino. This has led to studies on the methods to discover pure wino dark matter with a mass of less than 3 TeV \cite{Feng:1999fu,Gunion:1999jr, Ibe:2006de, Asai:2008sk, Abreu:1999qr, Buckley:2009kv}. Recent ATLAS and CMS searches \cite{Aad:2013yna,CMS:2014gxa} have excluded the pure wino LSP up to masses of 270 GeV and 260 GeV respectively using the disappearing track signature. The authors of \cite{Low:2014cba,Cirelli:2014dsa} show that the disappearing track search has a better chance of discovering pure wino dark matter than do the mono-jet, mono-photon, or vector boson fusion searches (all include large amounts of missing energy). In addition, they estimate that a future 100 TeV collider will be able to discover pure wino dark matter up to a mass of 3 TeV, necessary for generating the observed relic abundance. 

The pure wino dark matter of supersymmetric models has a direct counterpart in the minimal dark matter models \cite{Cirelli:2005uq,Cirelli:2007xd,Cirelli:2009uv}. Minimal dark matter models only add to the standard model particles under a representation of $SU(2)_L\otimes U(1)_Y$ which is automatically stable on the lifetime of the universe without resorting to symmetries beyond SM gauge interactions and Lorentz invariance.\footnote{The fermionic $SU(2)_L$ triplet (ie. wino) is not stable unless an addition symmetry is added. In this case, imposing the accidental $B-L$ symmetry of the SM allows for a stable dark matter candidate.} Two such models are cosmologically stable and avoid direct detection bounds: the scalar $SU(2)_L$ septet with hypercharge $Y=0$ and the fermionic $SU(2)_L$ quintuplet (fiveplet) with $Y=0$. The only models which pass direct detection bounds must have hypercharge $Y=0$. Representations smaller than the septet for scalars or the fiveplet for fermions allow for the neutral component to decay through Plank suppressed higher dimensional operators; representations larger than an octect for scalars or a fiveplet for fermions ruin perturbativity of $\alpha_2$ up to the Plank scale. As  scalar models generically contain scalar quartic couplings with the Higgs doublet, the fermionic model is usually referred to as `the' minimal dark matter candidate. While the previous studies have examined bounds from direct and indirect detection, there is no direct collider bound on the fermionic fiveplet. In this paper, we apply the 8 TeV ATLAS and CMS results to the fermionic model and exclude a fiveplet up to a mass of 267 GeV or 293 GeV depending on if the neutral component is Majorana or Dirac. We also extrapolate to the 14 TeV LHC and estimate that a Majorana (Dirac) fiveplet with a mass of 524 (599) GeV could be discovered at the end of the high luminosity run.

The rest of the paper is as follows. In the next section we review the set up of the minimal dark matter models. In section \ref{sec:search} we examine the current ATLAS and LHC searches for disappearing tracks and apply them to the minimal dark matter fiveplet. We extrapolate the results to $\sqrt{s}=14\tev$ in section \ref{sec:reach} to estimate the reach of the LHC for the model. We conclude in section \ref{sec:disc} and discuss how the doubly charged states present in the fiveplet model could be used to further increase the bounds in the future.

\section{Model}
\label{sec:model}

We examine the minimal dark matter model where $\chi$, a fermionic $SU(2)_L$ fiveplet with hypercharge $Y=0$, is added to the standard model. As a fiveplet, $\chi$ is represented as
\begin{equation}
\chi = \left( \chi^{++}, \chi^+, \chi^0, \chi^-, \chi^{--} \right).
\end{equation}
The lagrangian is then given by
\begin{equation}
\Lc = \Lc_{\text{SM}} + c~ \bar{\chi} \left(\imath \slashed{D} - M \right) \chi 
\end{equation}
where the constant c is 1/2 (1) if $\chi^0$ is a Majorana (Dirac) particle. In the covariant derivative, $\slashed{D}$, the $SU(2)_L$ generators give
\begin{eqnarray}
\left.t^a W^a_{\mu}\right|_{\text{fiveplet}} &=& \begin{pmatrix} 2W^3_{\mu} & \sqrt{2}W^+_{\mu} & 0 & 0 & 0 \\ \sqrt{2} W^-_{\mu} & W^3_{\mu} &\sqrt{3} W^+_{\mu} & 0 & 0 \\ 
0 &\sqrt{3} W^-_{\mu} & 0 & \sqrt{3} W^+_{\mu} & 0 \\
0 & 0 & \sqrt{3} W^-_{\mu} &- W^3_{\mu} & \sqrt{2} W^+_{\mu} \\
0 & 0 & 0& \sqrt{2} W^-_{\mu} & -2W^3_{\mu}\end{pmatrix}. \nonumber 
\end{eqnarray}

Conservation of Lorentz invariance and gauge symmetry protect the neutral component from decaying. The first operator which could lead to the fiveplet decay is dimension 6, given by 
\begin{equation}
\frac{c_2}{\Lambda^2} \chi L H H H.
\end{equation}
If $c_2$ is $\mathcal{O}(1)$, as long as $\Lambda \gtrsim 10^{14}\gev$, the lifetime of $\chi$ is greater than the age of the universe \cite{Cirelli:2005uq}. 

\begin{figure}[t]
\includegraphics[width=3in]{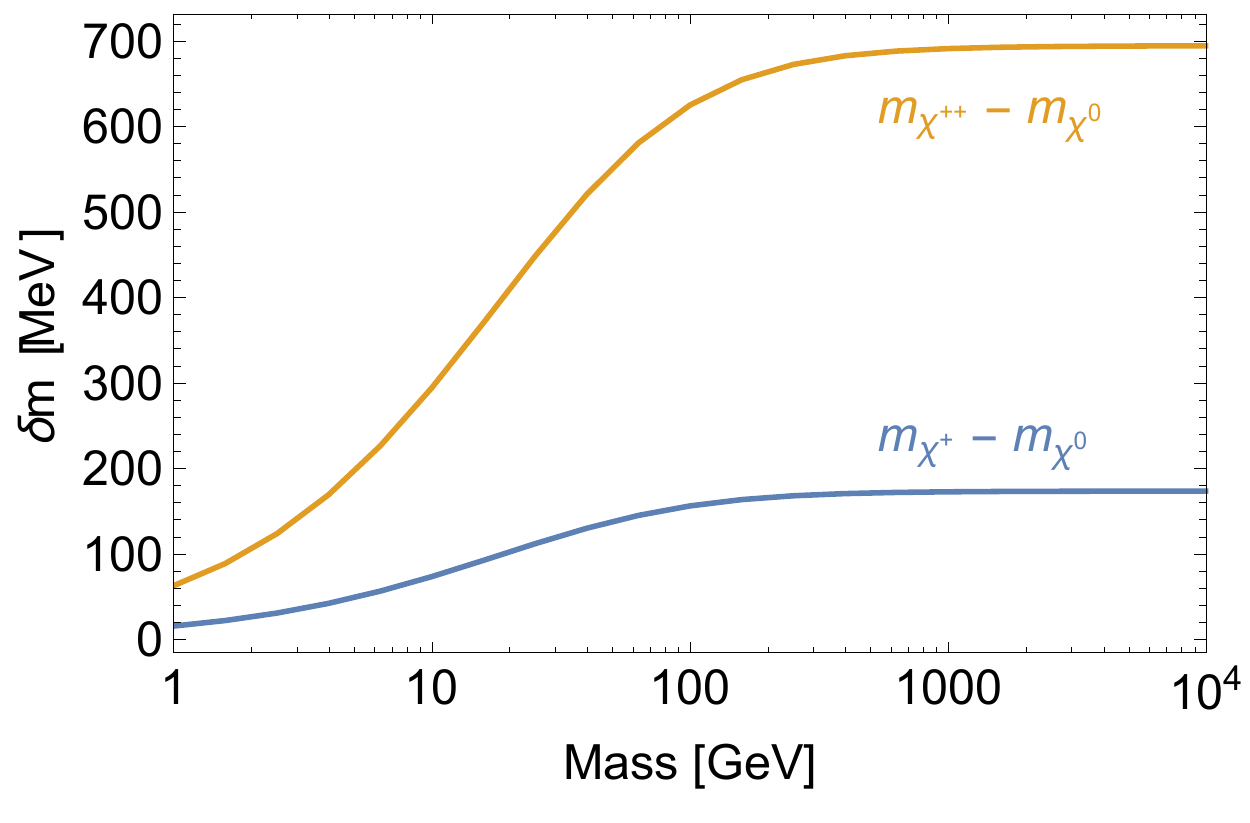}
\includegraphics[width=03in]{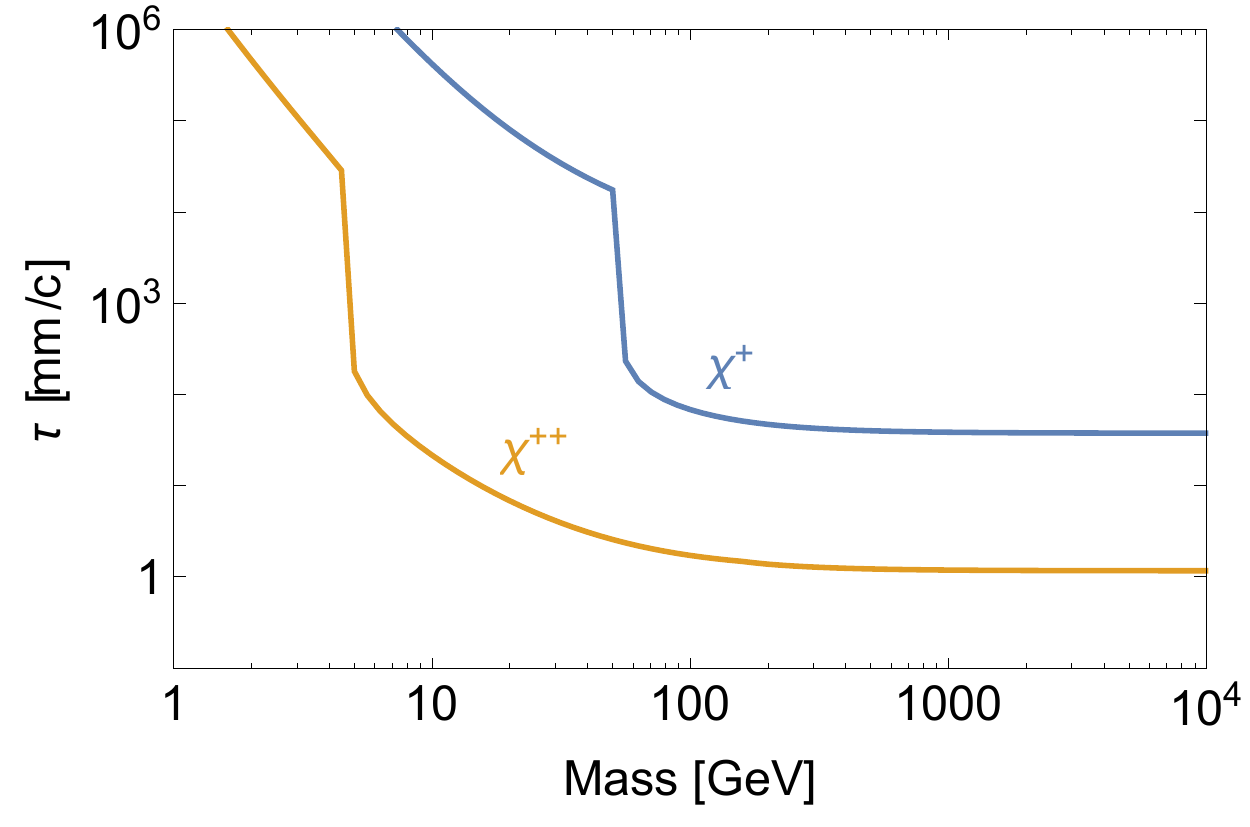}
\caption{Upper panel: Mass splitting between the charged states of the multiplet. Lower panel: Lifetime of the charged (blue) and doubly charged (orange) components of the fiveplet. The sharp drop corresponds to where the splitting is large enough to open the pion decay mode.}
\label{fig:MassAndLifetime}
\end{figure}

The entire multiplet has the same mass, $M$, at tree level, and this is the only free parameter as the only couplings are gauge interactions (ignoring higher-dimensional operators). At loop level, there are corrections which boost the mass of the charged components of the fiveplet with respect to the neutral component. We calculated this mass splitting by taking the difference of the 1 loop self energy diagrams for the different charged states. Analytic results for other multiplet representations can be found in \cite{Thomas:1998wy,Cirelli:2005uq,Cirelli:2009uv}. When $M >> m_W$, the singly charged component has a mass $\sim170$ MeV above the neutral component and the doubly charged state has a mass $\sim 690$ MeV above the neutral state. As the size of the mass splittings coming from the loop effects is small, it is reasonable to ask if higher dimensional operators can affect the mass splittings. At dimension 5 there are two operators,
\begin{equation}
\begin{aligned}
\mathcal{O}_1 &=\frac{c_{1,1}}{\Lambda}~ \bar{\chi}_{ijkl} \chi^{ijkl} H^m H^{\dagger}_m ~~~~\text{and}~~ \\
\mathcal{O}_2 &=\frac{c_{1,2}}{\Lambda}~ \bar{\chi}_{jklm} \chi^{iklm} H^j H^{\dagger}_i,
\end{aligned}
\end{equation}
where $ijklm$ are $SU(2)_L$ gauge indices. The first operator boosts the entire multiplet by the same amount, and serves only as a redefinition of $M$. The second operator boosts the charged components relative to the neutral component. If $\Lambda \gtrsim \mathcal{O}(10^8)\gev$, the contributions are less than an MeV. As we already need the dimension 6 operator to be suppressed more than this for $\chi$ to be stable, we will ignore the higher-dimension operators for the rest of this paper.

\begin{figure*}[t]
\begin{center}
\includegraphics[width=3in]{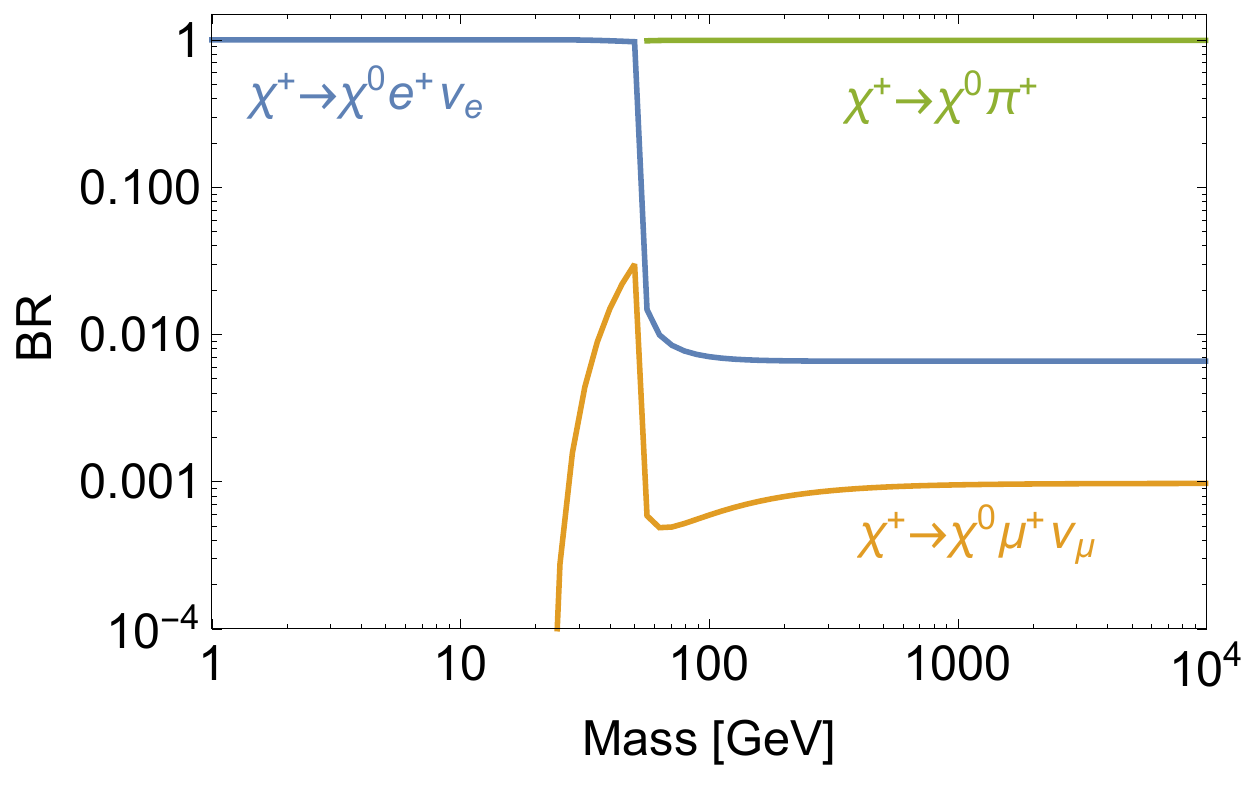}
\includegraphics[width=03in]{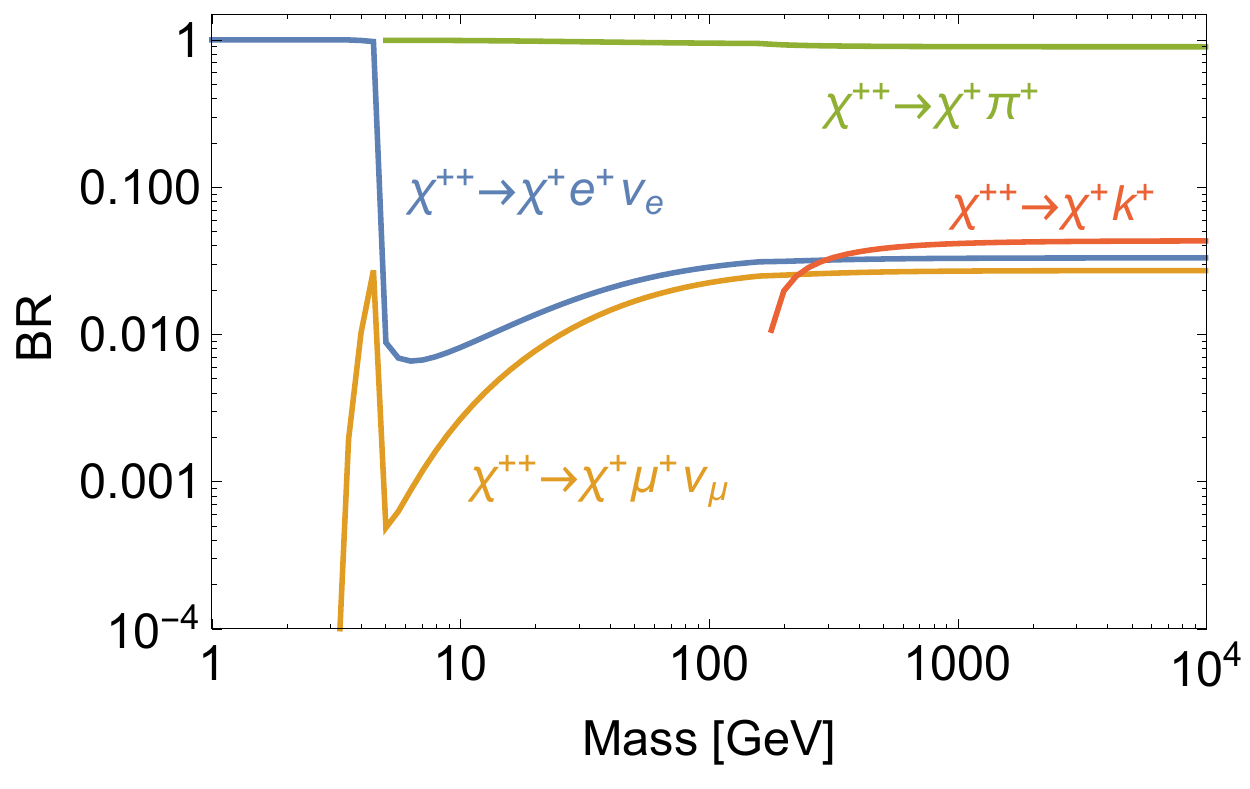}
\caption{Decay modes of the charged (left) and doubly charged (right) 5-plet states. The neutral state is stable.}
\label{fig:BR}
\end{center}
\end{figure*}

These small mass splittings leave little phase space available for decays. This leads to both the charged and the doubly charged states having finite lifetimes. Figure \ref{fig:MassAndLifetime} shows the mass splittings as a function of the multiplet mass and the associated lifetime expressed in units of mm/c.  The sharp drop in the lifetime happens when the mass splitting becomes large enough to decay through a charged pion.  

The branching ratios are shown in Fig.~\ref{fig:BR} with the singly charged on the left and the doubly charged on the right.  As the mass splitting is larger between the doubly charged and the singly charged states, the doubly charged fiveplet has an extra decay mode through the charged kaon, which opens around a mass of $\sim200\gev$. Despite this, both states are dominated by the charged pion channel for masses larger than $\sim100\gev$. It is worth noting that the lifetime of the singly charged fiveplet is smaller than the wino due to the group theory factors.

In reference \cite{Cirelli:2009uv} it was shown that the necessary mass of the fiveplet in order to achieve the correct relic abundance via thermal freeze out (including the Sommerfeld enhancement) is $\sim10$ TeV. This mass is too large to be probed by the LHC.\footnote{We are working on an exploration of the model at future colliders.} If the mass of the fiveplet is less than this then the amount dark matter from the thermally produced fiveplet is smaller than the observed abundance. While this does leave open the possibility that a light fiveplet provides some dark matter with another source of dark matter supplying the rest of the relic abundance, there are other options. One way to increase the relic abundance is through modified expansion history during freeze-out \cite{Salati:2002md}. There have also been studies on non-thermal production of WIMPs (usually the wino) \cite{Moroi:1999zb,Gelmini:2006pq,Acharya:2009zt,Watson:2009hw,Falkowski:2012fb,Baer:2014eja,Kane:2015qea} which use the late decay of a scalar field or moduli which either dumps more energy into the universe or decays directly to the dark matter. While we do not have an explicit model using a fiveplet, it is plausible that this could be done to have a light fiveplet give the correct relic abundance. It is therefore useful to see how LHC searches constrain the minimal dark matter model.

\section{Constraints from LHC8}
\label{sec:search}

We use both of the recent searches at the LHC for disappearing tracks to constrain the minimal dark matter fiveplet. We first present the search strategies used by ATLAS \cite{Aad:2013yna} and CMS \cite{CMS:2014gxa} and verify our results by comparing with the AMSB models used by the collaborations.

\subsection{ATLAS search strategy}

To start, we reproduce the benchmark of \cite{Aad:2013yna} using a pure wino of the MSSM with a mass splitting of $160$ MeV between the neutralino and chargino and a lifetime of $\tau_{\chi^{\pm}} = 0.2~\text{ns}$ for the chargino and a branching ratio of 100$\%$ for the $\chi^{\pm}\rightarrow \chi^0 \pi^{\pm}$ process. We simulate $\sqrt{s}=8$ TeV proton-proton collisions producing all possibilities of chargino/neutralino pairs plus up to two extra partonic jets using \textsc{Madgraph5@NLO} \cite{Alwall:2014hca}. At the partonic level, we demand that the leading jet has a transverse momentum of at least $70\gev$. The events are then showered, hadronized and matched using \textsc{Pythia6.4} \cite{Sjostrand:2006za} with the MLM matching scheme and passed through the \textsc{Delphes3} \cite{deFavereau:2013fsa} detector simulation using the default ATLAS card. Using \textsc{FastJet} \cite{Cacciari:2011ma}, the jets are clustered with the anti-k$_T$ algorithm \cite{Cacciari:2008gp} with a distance parameter of 0.4. Reconstructed jets are required to have $p_T > 20 \gev$ and $\left|\eta\right| < 2.8$, the electron candidates are required to have $p_T > 10\gev$ and $\left| \eta \right| < 2.47$, and the muon candidates are required to have $p_T > 10\gev$ and $\left| \eta \right| > 2.4$. These initial objects are then cleaned; any jet candidate within $\Delta R \equiv \sqrt{\left(\Delta \eta\right)^2 + \left (\Delta \phi\right)^2 } = 0.2$ of an electron is discarded. After that, any lepton within $\Delta R=0.4$ of a surviving jet is discarded.

The last object needed for the disappearing track signature is the track itself. \textsc{Delphes3} propagates final state particles and tracks final state charged particles with an $\eta$ and $p_T$ dependent efficiency. However, the disappearing chargino is not a final state particle so is not treated with this same method. Instead, we use the `truth' level information to get the chargino tracks in each event. As truth level objects, they are not simulated through a magnetic field. This should have little effect as the $p_T$ of the tracks will be required to be quite large, leading to almost straight tracks.

First, we find all isolated tracks in the event (including the charginos). To count as isolated, the sum of the transverse momentum of all other tracks within a cone of $\Delta R = 0.4$ about the track with a $p_T > 400\mev$ must be less than $4\%$ of the $p_T$ of the candidate isolated track. There must be no jets with $p_T > 45 \gev$ within $\Delta R = 0.4$ of the track. For the event to pass the selection criteria, the disappearing track (the chargino) must be the track with the largest $p_T$ (which also must be greater than 15 GeV) and lie within $0.1 < \left| \eta \right| < 1.9$.

A number of cuts are then used in the ATLAS analysis to separate the signals from the background.
\begin{itemize}
\item Leading jet $p_T > 90 \GeV$
\item Missing transverse energy $\slashed{E}_T > 90 \GeV$
\item $\Delta \phi^{\text{jet}-\slashed{E}_T}_{\text{min}} > 1.5$, where $\Delta \phi^{\text{jet}-\slashed{E}_T}_{\text{min}}$ is the azimuthal separation between the jet and the missing energy. If there are two or more jets with $p_T > 45 \GeV$, $\Delta \phi^{\text{jet}-\slashed{E}_T}_{\text{min}}$ is the smallest value from either of the two highest-$p_T$ jets. 
\item 30 cm $<$ transverse track length $<$ 80 cm
\end{itemize}

Before making a final cut on the transverse momentum of the track, ATLAS provides a benchmark for a chargino with a mass of $200\GeV$, in which with 20.3$\invfb$ of integrated luminosity, 18.4 Monte Carlo events pass the initial cuts. In our simulation, 23.9 events made it through the cuts. The ratio of these (.77) is taken as the efficiency of tracking the charginos within $0.1 < |\eta| < 1.9$, $p_T>15\gev$, and a transverse length between 30 and 80 cm.

Having defined our tracking efficiency, we scan over a mass range for the winos between 100 GeV and 500 GeV. The most constraining signal region defined in the ATLAS result is for $p_{T,\text{track}} > 200 \gev$, where the expected limit on the visible (after all cuts) cross section is $0.56^{+0.23}_{-0.16}$ fb and the observed limit is $0.44$ fb. For each mass point in our scan we compute the visible cross section defined as
\begin{equation}
\sigma_{\text{vis}} = \sigma_{\text{MC}} \times \epsilon_{\text{cuts}} \times \epsilon_{\text{tracking}}
\end{equation}
and compare to the ATLAS exclusions. The $95\%$ CL exclusion from ATLAS is 215-270 GeV while we obtain an exclusion between 204-243 GeV. Our results do not extend to as large of masses ($\sim10\%$ lower), which we take to imply that our results for the fiveplet will be conservative. Part of the reason for our differing results is the constant tracking efficiency we use, which in actuality should depend on the kinematics. We also used a fixed mass splitting and lifetime for all of the wino points, whereas they change with the mass of the wino in the ATLAS results. 

\subsection{CMS search strategy}

For the CMS search \cite{CMS:2014gxa}, we use the same \textsc{Madgraph} and \textsc{Pythia} events, but rerun them through \textsc{Delphes} using the default CMS card. The jets are clustered again using \textsc{FastJet} with the anti-k$_T$ algorithm with a distance parameter of 0.5. For an event to pass their basic selection requirements, it must have $\slashed{E}_T > 100\gev$ and at least one jet with $p_T>110\gev$ and $\left| \eta \right| < 2.4$. Additional jets with $p_T>30\gev$ and $\left| \eta\right| < 4.5$ are allowed provided that the azimuthal angle $\Delta\phi$ between any two jets is less than 2.5 radians and the minimum $\Delta\phi$ between the missing energy and either of the two highest-$p_T$ jets is greater than 0.5 radians.

This search does not veto events that have leptons, but instead vetos individual tracks that are within $\Delta R = 0.15$ of a lepton with a transverse momentum of at least 10 GeV. Other than this, the candidate tracks must have:
\begin{itemize}
	\item $p_T > 50\gev$ and $\left| \eta \right| < 2.1$
	\item No jet with $p_T > 30$ within $\Delta R= 0.3$ of the candidate track
	\item Sum of the transverse momentum of all other tracks within $\Delta R = 0.3$ of the candidate track must be less than $5\%$ of the candidate track $p_T$
\end{itemize}
To ensure that the potential disappearing track is well understood, tracks are vetoed if they are in regions of poor electron reconstruction ($1.42 < \left| \eta \right| < 1.65$) or poor muon reconstruction ($0.15 < \left| \eta \right| < 0.35$; $1.55 < \left|\eta \right| < 1.85$).

In the ATLAS analysis, we had to match to the number of Monte Carlo events to estimate the tracking efficiency for the charginos. However, CMS provides an efficiency of a track to pass the disappearing-track selection as a function of the transverse track length between 30-110 cm (table 8 of \cite{CMS:2014gxa}). We weight all of our events by this efficiency or when two charginos are present, by one minus the probability of missing both tracks.

After all cuts, the expected background is $1.4\pm1.2$ events with $19.5\invfb$ of integrated luminosity; 2 events are observed. Unfortunately, CMS does not then provide a cross section or number of signal events that are excluded at the $95\%$ CL.  We compute an approximate value using the expected number of events ($b_i$), the error in the SM expected number ($\sigma_{b,i}$), and the number of observed events ($n_i$). We plug these into the following equation \cite{6428calculation,4476calculation}, which is then numerically solved for the number of signal events $s_{i,95}$.
\begin{equation}
\frac{\int \delta b_i \text{Gaus} \left(\delta b_i, \frac{\sigma_{b,i}}{b_i} \right) \times \text{Pois} \left(n_i | b_i (b_i (1+\delta b_i)+ s_{i,95}\right)}{\int \delta b_i \text{Gaus}\left(\delta b_i, \frac{\sigma_{b,i}}{b_i}\right) \times \text{Pois}\left(n_i|b_i(1+\delta b_i) \right)} = 0.05
\label{eqn:si95}
\end{equation}
To compute the expected limit instead of the observed, we use $b_i$ in place of $n_i$. The Gaussian distribution is an attempt to take into account the uncertainties. With this, we obtain observed (expected) limits of 6.7 (5.6) events. We translate this into a limit on the visible cross section by dividing by the integrated luminosity. We note that doing this same process for the ATLAS search results in a limit that is weaker by $\sim30\%$ than what ATLAS reports. We therefore expect our results for the CMS search to be very conservative. In fact, our observed(expected) limits on the AMSB wino are 195(203) GeV whereas CMS reports a limit of 260 GeV at the $95\%$ CL. While our results for the CMS search are $20\%$ below the reported limits, they do at least agree with our ATLAS limits to within $10\%$.

\subsection{Applying the searches to the fiveplet}

\begin{figure*}[t]
\begin{center}
\includegraphics[width=3in]{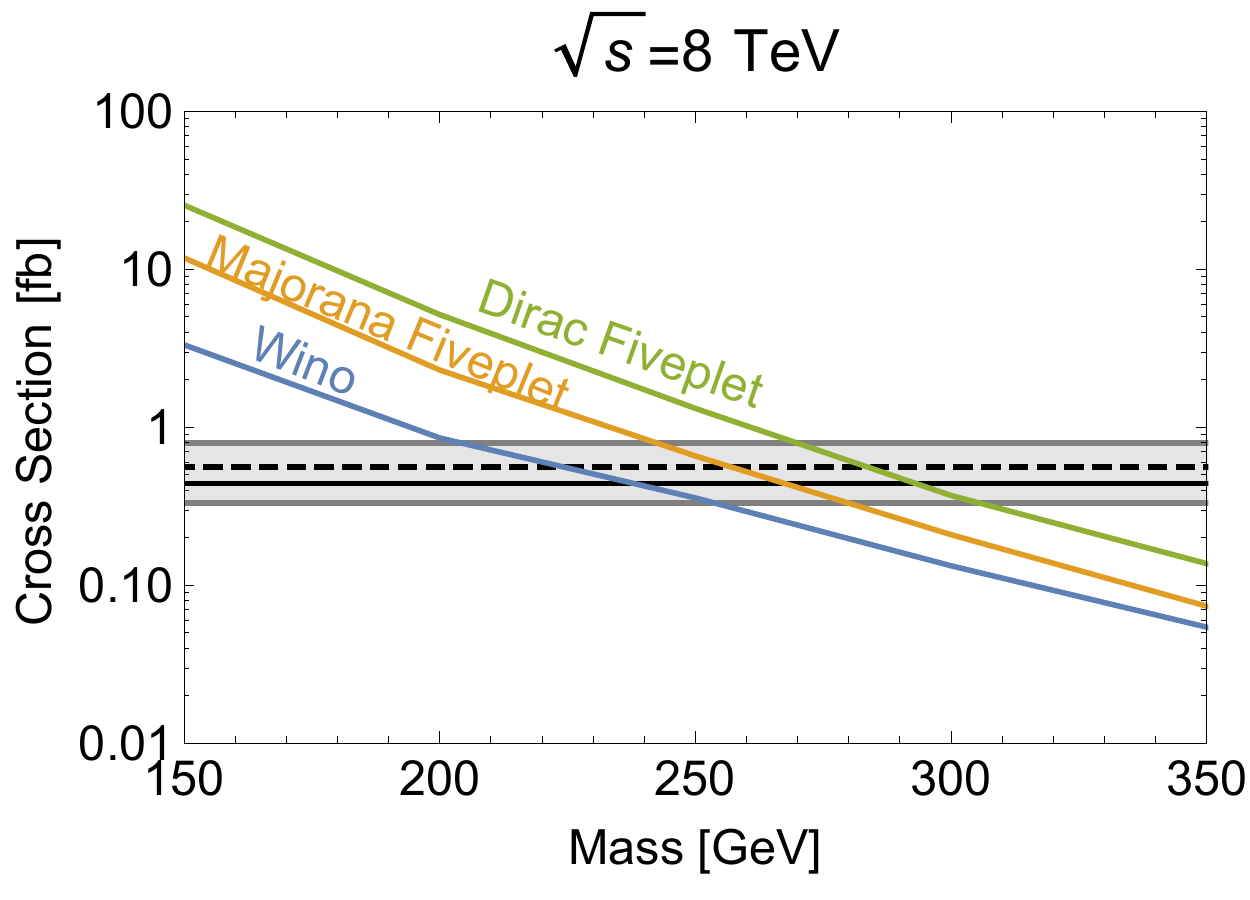}
\includegraphics[width=3in]{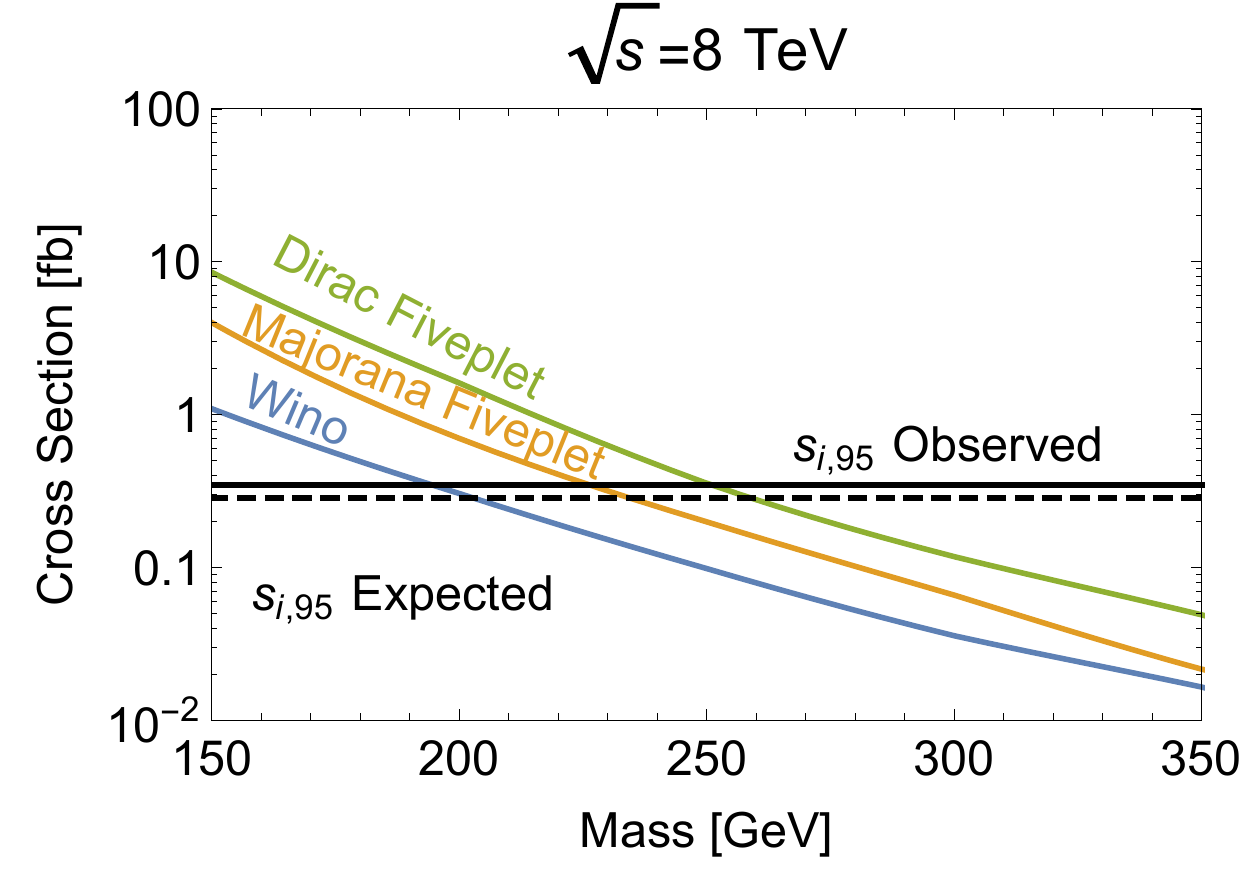}
\caption{Visible cross section after making search cuts used in the ATLAS (left)\cite{Aad:2013yna} and CMS (right)\cite{CMS:2014gxa} analyses. The grey shaded region along with the dashed line mark the expected $95\%$ CL exclusion with $1 \sigma$ error bands. The black line marks the observed exclusion. No values are given for the expected or observed $95\%$ CL exclusions for the CMS search, so a conservative estimate -- Eq.~\eqref{eqn:si95}-- is used. This is overly conservative, leading to the smaller exclusions. They are still within $10\%$ of the ATLAS results. }
\label{fig:ExlandReach}
\end{center}
\end{figure*}

To constrain the minimal dark matter fiveplet model, we use \textsc{FeynRules 2.0} \cite{Alloul:2013bka} to generate a UFO model file for \textsc{Madgraph}. The mass splittings, lifetimes, and branching ratios are different for each mass point, as shown in Figs.~\ref{fig:MassAndLifetime} and \ref{fig:BR}. We simulate events using the same generator level cuts as with the wino, and use modified \textsc{Delphes} cards to register the neutral fiveplet as part of the missing energy.

The group theory factors which make the decay length of the fiveplet smaller than the wino (and thus harder to travel far enough to be registered as a disappearing track) also increase the cross section. These two factors could essentially offset and lead to a fiveplet exclusion for the same mass as the wino. However, the fiveplet has doubly charged states not present in the wino model which could help extend the reach for the fiveplet model.

It is clear that the cross section for producing doubly charged states, either in pair production or along with a singly charged state, decays down to a singly charged state should add to the overall signature of the disappearing track. However, it is more subtle than that. In order to have a well reconstructed track, the ATLAS (CMS) analysis demand the impact parameter for the disappearing track be $|d_0| < 0.1\,(0.2)\units{mm}$ and $|z_0 \sin\theta | < 0.5\,(5.0)\units{mm}$ where $d_0$ and $z_0$ are the transverse and longitudinal impact parameters respectively. Figure \ref{fig:MassAndLifetime} shows that the doubly charged particle can easily travel on the order of mm in the detector. This has the potential to then lead to a large impact parameter and therefore get missed in the event selection. We do not find that this is an issue for the following reasons.
\begin{itemize}
	\item To conserve momentum in the decay of the doubly charged particle (at rest), the heavy (compared to the pion) charged fiveplet does not get much velocity. When this is then boosted to the lab frame, the doubly charged and the singly charged fiveplets travel in almost the same direction. In our simulations the angle between the velocities was $< 0.01$ radians.
	\item The search demands a large $p_T$ of the track. This leads to a minuscule change in the radius of curvature of the charged and doubly charged particles in the magnetic field. We examined this analytically by letting a doubly charged particle propagate through a magnetic field until it decayed. The last bullet point showed that the singly charged particle has nearly the same momentum at this point. We then trace back from the decay point using a singly charged particle and find that it is within micrometers of the origin of the doubly charged particle.
\end{itemize}
Applying these two statements implies that although there is a `displaced' vertex where the doubly charged particle decays, there will be no indication that this happened. The singly charged plus doubly charged track will be reconstructed as one track coming from the primary vertex. With this assumption, to find the track length, we get the location where the neutral state is created, and do not worry about the particle's history. 

Using the  ATLAS search we obtain expected 95$\%$ exclusions of  242-271 GeV and 269-296 GeV depending on whether $\chi^0$ is Majorana or Dirac respectively. With CMS search, we obtain expected (observed) exclusion of 234 (226) and 259 (251) GeV for the Majorana and Dirac fiveplets respectively. We again emphasize that our low exclusions for the CMS search come from the conservative estimate of the number of signal events that can be excluded at the $95\%$ CL. The exclusion limits for the wino and the two fiveplet models are shown in Fig.~\ref{fig:ExlandReach} where the grey region is the expected excluded cross section and the solid line is the observed exclusion.

\section{LHC14 Reach}
\label{sec:reach}
The backgrounds for the disappearing track signature involve charged hadrons interacting with large momentum exchange within the detectors. This  is therefore hard to simulate and extrapolate to the 14 TeV LHC run.  ATLAS fits the shape of the $p_T$-mismeasured tracks as $d\sigma/d p_{T}^{\text{track}} = \left(p_{T}^{\text{track}}\right)^{-a}$ where $a = 1.78\pm0.05$, and they note that this background is the dominant background at large $p_{T}^{\text{track}}$. Following the example of \cite{Low:2014cba} and \cite{Cirelli:2014dsa}, we first normalize to the expected background events in \cite{Aad:2013yna} at $\sqrt{s}=8\tev$ with $p_T^{\text{track}} > 200\gev$. This normalization of the shape is then extrapolated to 14 TeV by taking the ratio of $Z(\nu\bar{\nu}) +\text{jets}$ cross section passing initial analysis cuts on the jet $p_T$, $\slashed{E}_T$, and $\Delta \phi_{\text{min}}^{\text{jet}-\slashed{E}_T}$ at $\sqrt{s}=8$ and 14 TeV.

It is not clear that the background should keep the same shape with a change in collider energy or the $p_T$ of the jet. To account for this uncertainty the authors of \cite{Low:2014cba} and \cite{Cirelli:2014dsa} examine the significance in two extremes, where the background is larger or smaller by a factor of 5. In  \cite{Cirelli:2014dsa}, the authors also allowed the index of $a$ to vary up to $5-\sigma$ from its central value and found that this band was smaller than that of using $20\%-500\%$ background. We will use the same conventions and present all results in terms of this $20\%-500\%$ background uncertainty. However, we note that this could greatly be improved with a study of the $p_T$ -mismeasured tracks at the 13 and 14 TeV runs of the LHC.

We use the 14 TeV analysis cuts presented in \cite{Cirelli:2014dsa}, which were optimized over the $\slashed{E}_T$, $p_T^{\text{track}}$ and $p_T (j_1)$ for pure wino dark matter. These are summarized by
\begin{equation}
\slashed{E}_T > 220 \gev,~~p_T(j_1) > 220 \gev,~~p_T^{\text{track}} > 320\gev.
\end{equation}
They also include extra jets with $p_T>70\gev$, again keeping $\phi_{\text{min}}^{\text{jet}-\slashed{E}_T} > 1.5$ for the hardest two jets.

\begin{figure}[t]
\begin{center}
\includegraphics[width=3in]{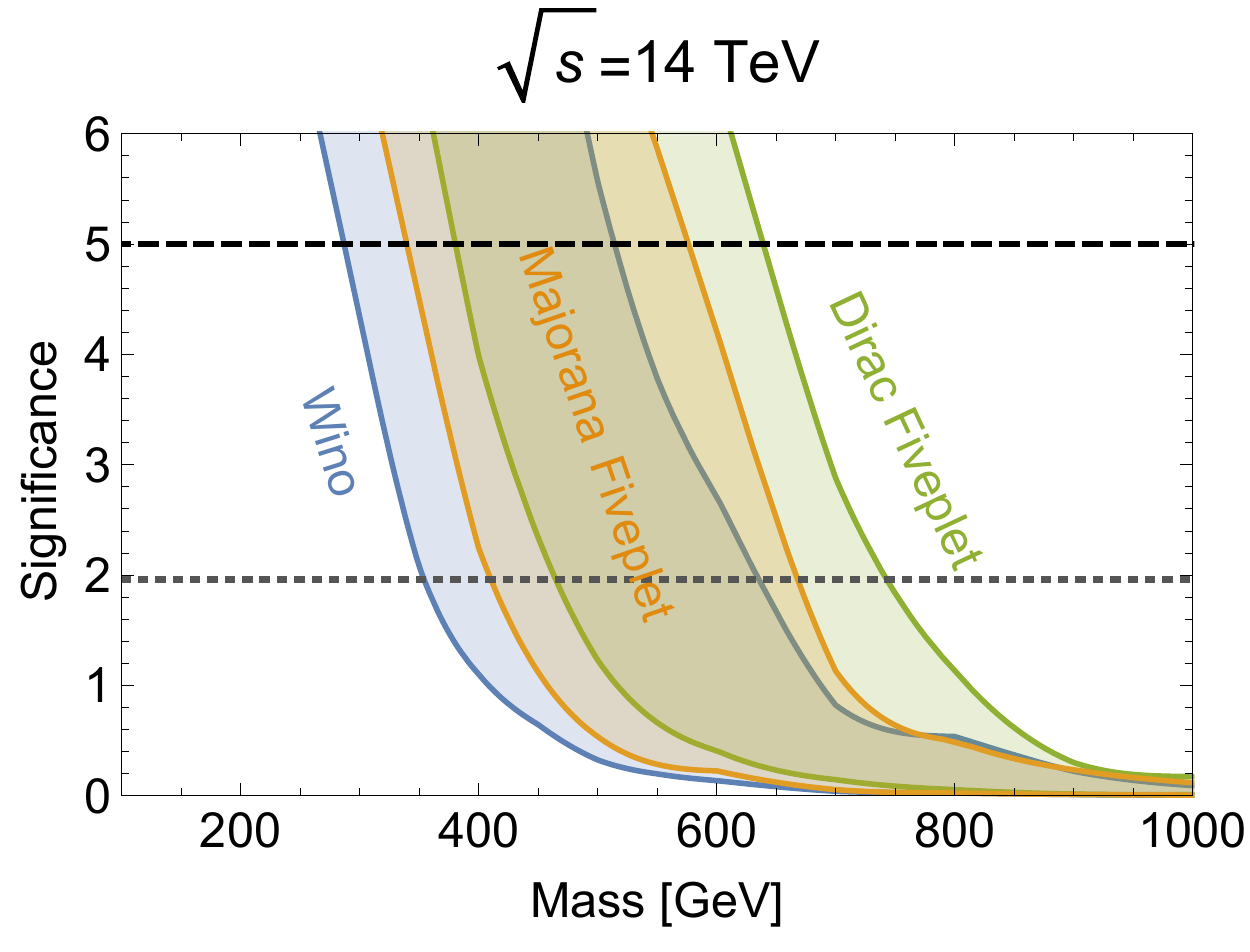}
\caption{Reach for the $\sqrt{s}=14\tev$ LHC with $3\invab$ of data. The bands are generated by varying the background between $20\%$ and $500\%$ to account for the uncertainty in the extrapolation between 8 and 14 TeV. }
\label{fig:Reach}
\end{center}
\end{figure}

Our results for the LHC reach are presented in Fig.~\ref{fig:Reach}. We change the significance measure from that of Eqn.~\eqref{eqn:si95} to allow for both a $5\sigma$ discovery and a $95\%$ CL exclusion. The $y$-axis is the significance given $3 \invab$ of integrated luminosty, defined as 
\begin{equation}
\text{Significance} = \frac{S}{\sqrt{B+\alpha^2 B^2 + \beta^2 S^2}},
\end{equation}
where $S$ and $B$ are the number of signal and background events passing the cuts.  The systematic uncertainties are denoted as $\alpha$ for the background and $\beta$ for the signal and are conservatively given the values of $\alpha = 20\%$ and $\beta=10\%$ as done in \cite{Low:2014cba,Cirelli:2014dsa}. The black dashed line shows when the significance reaches $5\sigma$, denoting the possibility of discovery. Similarly, the dotted line represents the $1.96\sigma$ level, which then corresponds to what could be excluded at the $95\%$ CL. The results are summarized in Tab.~\ref{tab:results}. 

Our 14 TeV results for the wino are consistent with \cite{Cirelli:2014dsa} and are able to exclude to a larger mass than \cite{Low:2014cba} due to the optimized cuts. We have not tried to optimize specifically for the fiveplet models, but the optimal cuts should be similar since the lifetime and decay modes are so similar. The presence of the extra, doubly charged particles should not have an effect on the track $p_T$.

With the wino optimized cuts, a Majorana fiveplet could be discovered with a mass between 340-580 GeV or excluded between 410-670 GeV. If the fiveplet is Dirac, then it could be discovered with a mass between 380-640 GeV or excluded between 465-745 GeV. These are about 10-20$\%$ better than the wino.

\begin{table*}[t]
\begin{tabular}{c| c c | c c | c c c | c c c }
\hline\hline
\multirow{3}{*}{Model}& \multicolumn{4}{c|}{$\sqrt{s}=8\tev$} & \multicolumn{6}{c}{$\sqrt{s}=14\tev$} \\
\cline{2-11}
& \multicolumn{2}{c|}{ATLAS} & \multicolumn{2}{c|}{CMS} & \multicolumn{3}{c|}{Exclude} & \multicolumn{3}{c}{Discover} \\
&Expected& Observed & Expected&Observed & $500\%$ & $100\%$ & $20\%$ & $500\%$ & $100\%$ & $20\%$ \\
\hline
Wino & 224 &238 & 203 & 195 & 354 & 483 & 635 & 287 & 394 & 514  \\
Majorana Fiveplet & 256 & 267 & 234 & 226 & 410& 524& 668 & 340 &448 &576  \\
Dirac Fiveplet & 283  & 293 & 259 & 251 & 465 & 599 & 743 & 381 & 503 & 639  \\
\hline
\hline
\end{tabular}
\caption{Summary of current bound and future reach at the LHC for the minimal dark matter fiveplet. The numbers are the masses in units of GeV. The 14 TeV numbers show the masses that could be excluded or discovered with $3\invab$ of data. The 500$\%$, 100$\%$, and $20\%$ columns labels refer to the amount of background compared to the extrapolation.}
\label{tab:results}
\end{table*}

\section{Discussion and Conclusions}
\label{sec:disc}

Minimal dark matter models add a single electroweak multiplet to the standard model particle content. When the representation is large, the mass splittings between different states in the multiplet only come in at loop level. The size of the representation under $SU(2)$ can automatically make the lightest member of the multiplet stable. The small mass splittings between the states of the multiplet leave little phase space for decays, leading to charged tracks which travel a finite distance in the detectors. If the charged particle decays in the detector volume, the track appears to disappear before reaching the calorimeter. 

ATLAS and CMS have searched for such disappearing tracks, interpreted in the framework of the anomaly mediated supersymmetry breaking scenarios where the wino is the LSP. We duplicated the search strategies and found that the $SU(2)_L$ fermion fiveplet can be excluded to masses $\sim10(20)\%$ larger than the wino when the fiveplet is real(complex). The limits on the fiveplet models are larger in part due to the increased cross section coming from group theory factors in the coupling of the $W$ to the multiplet. In addition, the fiveplet models contain a doubly charged particle. In the analysis presented here, the presence of these extra, doubly charged particles is only used to enhance the cross section, but there is no other attempt to identify or use them.

While the boosted cross section from the doubly charged particles is nice, we believe that more could be done with them. For instance, if an excess is observed in the next run of the LHC in the disappearing track signature, an attempt to identify doubly charged particles could help distinguish a triplet (wino) from a fiveplet. In the wino model, the largest production process is $\chi^{\pm}\chi^0$, and $\sim 70\%$ of the disappearing track events contained a single chargino. Meanwhile, the fiveplet model adds the process of $\chi^{\pm\pm} \chi^{\mp}$ and $\chi^{\pm\pm} \chi^{\mp\mp}$ which both yield events with two charged tracks. In our simulation only $\sim30\%$ of the disappearing track events of the fiveplet model contained only one singly charged particle in the event. 

The current strategies used by the collaborations are aimed at discovering a single track. In the ATLAS search, the disappearing track must be the track with the largest $p_T$. Similarly, the CMS cuts, particularly the wide range of $|\eta|$ which is vetoed for poor lepton reconstruction, keeps the count rate very low (only 2 observed events in the signal region, with a background of 1.4). Having two tracks fall in the allowed region will be even harder. To travel the transverse distance needed to be registered as disappearing, the particles (especially the fiveplet with the shorter lifetime) must be quite boosted and in the central part of the detector. Usually one of the pair produced particles satisfies this, while the other is either not boosted as much or not as central. Widening the range of $|\eta|$ and decreasing the minimum transverse length (only for extra tracks) could aid in the detection of a second, short track. We are also working on a study to larger center of mass energy, such as a $100\tev$ collider, where both particles could receive a large enough boost to be registered as disappearing.  A lack of a signal for two disappearing tracks could then greatly exclude the fiveplet model where most events contain two charged tracks. This search method could extend to any model with pair production of a charged particle with a lifetime on the order of mm/c. 

We also note that in the fiveplet model, the doubly charged particle still travels a finite distance in the detector. While this does not lead to a observable displaced track, it could allow the soft pion from the decay reach the inner detector. In such a situation, the disappearing track would have a pion with $\sim 1 \GeV$ of energy looping around at the start and the end of the track. This is distinct from the wino where the pion only occurs at the end of the track.

\acknowledgements
We thank Tilman Plehn for discussions which got this project started and Adam Martin and Joe Bramante for many useful discussions which helped it progress. We are also grateful for the comments on the draft received from Adam Martin, Antonio Delgado, and Landon Lehmen. This work was supported in part by the National Science Foundation under grant PHY-1215979.

\bibliography{FivepletBib}

\end{document}